\documentstyle[aps]{revtex}
\begin {document}
\title {Slow dynamics of Ising models with energy barriers}
\author{A.~Lipowski$^{1)}$, D.~Johnston$^{2)}$ and D.~Espriu$^{3)}$}
\address{
$^{1)}$ Department of Physics, A.~Mickiewicz University,
61-614 Pozna\'{n}, Poland\\
$^{2)}$ Department of Mathematics, Heriot-Watt University
EH14 4AS Edinburgh, United Kingdom\\
$^{3)}$ Department of Physics, University of Barcelona, 08028 Barcelona, Spain}
\maketitle
\begin {abstract}
Using Monte Carlo simulations we study the dynamics of three-dimensional
Ising models with
nearest-, next-nearest-, and four-spin (plaquette) interactions.
During coarsening, such models develop growing energy barriers, which leads 
to very slow dynamics at
low temperature.
As already reported, the model with only the plaquette interaction exhibits
some of the features characteristic of ordinary glasses:
strong metastability of the supercooled liquid, a weak increase of the 
characteristic length under cooling, stretched-exponential relaxation and 
aging.
The addition of two-spin interactions, in general, destroys such  behaviour: the liquid
phase loses metastability and the slow-dynamics regime terminates well below
the melting transition, which is presumably related with a certain 
corner-rounding transition.
However, for a particular choice of interaction constants, when the ground 
state is strongly degenerate, our simulations suggest that the slow-dynamics
regime extends up to the melting transition.
The analysis of these models leads us to the conjecture that in the four-spin Ising 
model domain walls lose their tension at the glassy transition and 
that they are basically tensionless in the glassy phase.
\end {abstract}
\section{Introduction}
A lot of efforts have been devoted in the last twenty years to understanding 
the behaviour of various glassy and disordered systems~\cite{GOTZE}.
Such systems, which include conventional glasses, spin glasses, amorphous 
semiconductors, and many others are of great importance both experimental 
and theoretical.
However, despite intensive research, our understanding of such systems is 
still limited.
For example, even the very nature of the glassy phase in spin-glasses is 
still a very controversial issue~\cite{PARISI,FISHER,HOUDAYER}.
Although they are much more abundant, conventional glasses seem to pose even 
a greater puzzle.
Why do supercooled liquids fall out of equilibrium at a more or less well 
defined temperature?
Why do they collapse into the glassy state when the cooling is fast enough 
and into the crystalline phase when the cooling is slow?
These fundamental questions still await definitive answers.
One of the important problems in physics of conventional glasses is the
continuing
lack of a satisfactory microscopic model of such systems.
In this respect the situation is much better for spin-glasses where it is 
commonly accepted that models containing quenched disorder 
correctly describe physics of such systems.
Lattice realizations of such models are a particularly valuable source of 
information about spin-glasses~\cite{YOUNG}.
The most realistic models of conventional glasses, so-called off-lattice 
models, still constitute an enormous computational challenge although 
progress in this field is also being made~\cite{KOB}.

A model of conventional glasses should be capable of describing (at least) 
three phases: liquid, glass and crystal.
The actual state of the system should be determined by control 
parameter(s) (e.g., temperature) and possibly also its history.
Since the glass is regarded as a liquid trapped during the 
falling out of equilibrium, the model should possess such a trapping mechanism.
In spin glasses the trapping mechanism is related with energy barriers 
generated by quenched disorder~\cite{FISHER}.
On the other hand we do not expect the quenched disorder to be a relevant 
factor in conventional glasses because models with strong quenched disorder 
are unlikely to exhibit periodic solutions (which are needed for the model 
to be in the crystal phase).
Recently, various lattice models, which do not contain quenched disorder, 
were studied which have some features of conventional glasses.
Some of these models are infinite-dimensional and their 
thermodynamical properties 
can be found exactly~\cite{BERNASCONI}.
There are also finite dimensional models whose dynamics exhibit some glassy
behaviour~\cite{SHORE,RIEGER}

Recently, it has been shown that the three dimensional Ising model with the 
four-spin (plaquette) interaction also exhibits some glassy 
features~\cite{LIP,LIPDES,LIPDES2}.
This model undergoes a first-order phase transition between low-temperature 
(crystal) and high-temperature (liquid) phases.
However, when conventional simulation techniques are used,  the transition is 
screened by a very strong metastability during heating as well as cooling.
For temperatures lower than the limit of metastability of the liquid phase, 
the model has a very slow coarsening dynamics.
In addition, the zero-temperature characteristic length increases very slowly
as a function of the inverse cooling rate, which is also an expected property
of glasses.
Further evidence of the glassy behaviour in this model has been recently reported 
by Swift {\it et al.}~\cite{SWIFT}.
They have shown that the glassy transition coincides with the 
divergence of a certain relaxation time and that aging properties of the model 
are also typical of glassy systems.
They have also observed that some time correlation functions may decay as 
stretched exponentials.

These results strongly suggest that the model with four-spin interactions 
might describe important aspects of the glassy transition.
It would be interesting to find which properties of this model are responsible
for such a behaviour.
It has been already suggested~\cite{LIP,LIPDES,LIPDES2} that the trapping mechanism
might be related with diverging energy barriers.
These barriers would arise in this model basically due to the same mechanism 
as in a model 
with competing nearest-neighbour and next-nearest-neighbour interactions examined by Shore {\it et al.}~\cite{SHORE}
(the SS model for short).
However, the behaviour of the SS model is not fully consistent with our 
conception of glasses since it orders too quickly under cooling~\cite{SHORE}.
It was also suggested that the difference in the behaviour of the 
SS and four-spin models
might be related with the degeneracy of the ground state in the four-spin model.
This degeneracy might lead to some entropy barriers, which would be responsible
for the strong metastability of the liquid phase.

In the present paper, using Monte Carlo simulations, we examine a certain class 
of three-dimensional Ising models which generate energy barriers.
These models are described by the following Hamiltonian
\begin{equation}
H =-J_1\sum_{<i,j>} S_iS_j -J_2\sum_{<<i,j>>} S_iS_j -
J_4\sum_{[i,j,k,l]} S_iS_jS_kS_l.
\label {1}
\end{equation}
In the above expression $<..>$ and $<<..>>$ denote pairs of nearest and next-nearest neighbours, respectively, and $[i,j,k,l]$ stands for summation over 
elementary plaquettes.
In general, these models have double degenerate ground state and our 
simulations suggest that the dynamical properties in 
this case are similar to the SS model.
However, when the interaction constants are such that the model has a strongly
degenerate ground state (gonihedric case), the dynamical properties change.
Simulations suggest that two types of dynamical behaviour appear.
In the first type the model behaves similarly to the already described 
four-spin model.
In the second type, the glassy transition appears to coincide with the thermodynamic transition.

Such behaviour gives rise to the following questions: why 
a glassy transition appears 
only in certain systems with slow dynamics and what is its 
nature.
Analysis of the ground-state structure and thermodynamic properties of 
models studied here prompts the following conjecture, which, if
confirmed, would constitute an important result of the present paper: 
at the glassy transition the domain walls
lose their surface tension, and, as a result, the glassy phase is composed 
of tensionless domains.
Although based on the analysis of Ising models, we hope that such an 
interpretation might shed some light on the nature of the glassy transition 
in more realistic systems also.
Moreover, such an interpretation of glassy phase is in accord with 
some recent hypotheses concerning the nature of the glassy phase in spin 
glasses~\cite{HOUDAYER}.

In section II we discuss briefly the properties of the four-spin model.
In section III we present the results of our simulations for the gonihedric 
case.
The case of the doubly-degenerate ground state is discussed in section IV.
Section V contains a summary of our results and some arguments on the
nature of the glassy transition.
\section{Ising model with plaquette interactions}
This model corresponds to the case $J_1=J_2=0,J_4=1$, and has been already studied 
using Cluster Variational Method~\cite{CIRILLO} and Monte Carlo 
simulations~\cite{BAIG,LIP,LIPDES,LIPDES2,SWIFT}.

Clearly, the ferromagnetic configuration is a ground-state configuration of this model.
It is also easy to realize that flipping coplanar spins does not change the energy.
Thus any configuration obtained from the ferromagnetic configuration by flipping coplanar 
spins is also a ground-state configuration.
Moreover any combination of such coplanar flippings (even for crossing 
planes) does not increase the energy.
Simple analysis along these lines shows that for the model on the lattice of the linear size
$L$ the degeneracy of the ground state is equal to $2^{3L}$.
Although ground state of this model is strongly degenerate its ground-state entropy is zero.

The model undergoes a first-order thermodynamic transition at $T=T_{{\rm c}}\sim 3.6$ which is, however, screened by very strong metastability~\cite{LIPDES}.
As a result, when heated or cooled, the transition observed in simulations is 
shifted to $T\sim 3.9$ or $T\sim 3.4$, respectively.
Transitions at these spinodals are accompanied by peaks in the specific heat.

The low-temperature spinodal $T\sim 3.4$ seems to coincide with the glassy transition.
Below this temperature the model exhibits very slow coarsening dynamics~\cite{LIPDES} as well as aging properties which are characteristic of glassy systems~\cite{SWIFT}.
A certain characteristic time, which governs the relaxation of energy-energy correlation functions, also seems to diverge at this temperature~\cite{SWIFT}.
In addition, the behaviour of the model under continuous cooling supports
the glassy-transition interpretation of this temperature~\cite{LIPDES2}.
\section{Gonihedric Ising model}
\subsection{Ground state and thermodynamics}
It has already been suggested that the slow dynamics of 
the four-spin model might be 
related with energy barriers generated in that model~\cite{LIP,LIPDES}.
These barriers arise due to the shape-dependence of the energy 
of excitations: it is 
not only the size of an excitation which determines its energy but also its 
shape.
Such shape dependence appears also in the SS model.

Are there any other models which could have a similar property?
In our opinion, the shape dependence of energy of excitations should be 
rather a robust feature of Ising-type models.
It is only in some specific cases, like the standard nearest-neighbour case, when this 
energy does {\it not} depend on the shape of an excitation.
In particular, energy barriers appear in model (\ref{1}).
The Hamiltonian of this model is quite general and it includes both the 
four-spin model and the Shore {\it et al.}'s model ($J_4=0, J_1>0, J_2<0$).

In the present section we examine a class of models described by this
Hamiltonian, namely gonihedric models~\cite{SAVVIDY,SAV}.
These models correspond to the following choice of interaction constants: 
$J_1=2k, J_2=-\frac{k}{2}$ and $J_4=\frac{1}{2}(k-1)$.
Gonihedric models have a strongly degenerate ground state.
In addition to the ferromagnetic ground state any configuration obtained by 
flipping coplanar spins also minimizes the Hamiltonian.
Any  combination of such flips does not increases the energy, provided 
that flipping planes do not cross.
As a particular example of such a ground-state we can mention lamellar 
configurations where e.g., every second plane of spins is flipped.
Although lamellar structures constitute a legitimate ground-state, they do not
survive at finite temperature as shown by Cirillo {\it et al.} using the 
Cluster Variational Method~\cite{CIRILLO}.
We will return to this feature in the last section.

For $k=0$ the gonihedric model is equivalent to the four-spin Ising model.
In this case the model has an additional symmetry which implies a larger 
degeneracy of the ground state
since the flipping planes can now cross.
As a result we obtain that antiferromagnetic configurations belong to 
the ground state.
Further analysis of differences between the $k=0$ and the $k\neq 0$ cases is 
postponed to the last section.

Gonihedric models are expected to undergo a thermodynamic transition which for 
$k<k_{{\rm tr}}$ is of first order and for $k>k_{{\rm tr}}$ is of second order.
Only very rough estimations of $k_{{\rm tr}} (\sim 0.5)$ are 
known~\cite{CIRILLO}.
In this section we analyze dynamical properties of the gonihedric model for 
$k=2$, i.e., for a value with a continuous transition.
Our results were obtained using a standard Monte Carlo method with random 
sequential update using Metropolis algorithm~\cite{BINDER}.
Some details can be found elsewhere~\cite{LIPDES,LIPDES2}.

To find the thermodynamic transitions we measured the specific heat.
Our simulations, which were made for various linear sizes $L$ up to $L=40$, 
locate the peak at $T_{{\rm c}}\sim 2.35$, which is a good agreement with the 
Cluster Variational Method estimation~\cite{CIRILLO}.
The absence of hysteresis effects confirms that the transition at 
$T=T_{{\rm c}}$ is continuous. (Although the nature of the thermodynamic 
transition in gonihedric models is an interesting and still open problem, its
further analysis is not an objective of the present paper).
\subsection{Dynamics}
An important indication of glassy dynamics is a slow evolution of a random
quench.
For usual models with nonconservative dynamics one expects~\cite{BRAY} that 
the characteristic length $l$ increases with time $t$ as $l\sim t^{1/2}$.
However, in glassy systems $l$ should increase much more slowly and presumably only
logarithmically with time ($l\sim \ln t$).
Such a behaviour most likely appears in the SS model and in 
the four-spin model.
In the following we present the results of our simulations of the evolution 
of quenches in the $k=2$ model.
We measured the excess energy $\delta E=E-E_{\infty}$, where $E_{\infty}$ is 
the equilibrium energy.
Our results for temperatures $T<T_{{\rm c}}$ are shown in Fig.~\ref{f1}.
To relate the characteristic length with the energy excess we can employ the 
frequently used relation~\cite{SHORE,LIPDES}
\begin{equation}
l \sim 1/\delta E.
\label{2}
\end{equation}
With this identification from Fig.~\ref{f1} we infer that for all the examined 
temperatures the asymptotic increase of $l$ is much slower than $t^{1/2}$.
At the end of this section we will argue that the relation (\ref{2}) most 
likely does not hold for this model and Fig.~\ref{f1} actually suggest that
the increase of $l$ is slower than $t^{1/4}$.
Since there are no theoretical arguments for such a slow algebraic increase in
our opinion it is quite plausible that asymptotically we have $l\sim \ln t$.
Such a slow increase of $l$ is most likely due to energy barriers.

An alternative technique to examine the dynamics of the model is to measure 
characteristic times of certain processes.
For example, one can measure the average time $\tau$ needed for the inversion 
of a cubic like excitation.
In this method, which was already applied to similar 
models~\cite{SHORE,LIPDES}, one prepares the system of the size $L$ with fixed 
boundary conditions and interior spins which are opposite to the boundary 
spins.
One expects that after some time, the system will invert the interior spins.
For a two-spin Ising model or SS model above the corner 
rounding transition, $\tau \sim L^2$, which indicates a relatively fast 
dynamics (naive inversion of this relation gives $l\sim t^{1/2}$).
On the other hand, in the SS model below the corner rounding transition and in 
the four-spin model $\tau$ increases much faster, presumably exponentially,
with $L$.

We measured the time needed for magnetization of the interior spins to reach 
the equilibrium value at a given temperature and the results are shown in
Fig.~\ref{f2}.
These results show that $\tau$ even at the highest examined temperature 
increases faster than $L^2$.
This is a potential indication of an exponential increase $\tau \sim a^L (a>1)$ in 
the entire low-temperature phase.

Additional confirmation of such behaviour is obtained from simulations of 
this model under the continuous cooling.
Similarly to simulations of the four-spin model~\cite{LIPDES2}, we relax the 
random sample at a temperature $T_0>T_{{\rm c}}$ and then 
continuously lower the temperature according to the formula $T(t)=T_0-rt$, 
where $r$ is the cooling rate.
When the temperature is reduced below the critical point $T_{{\rm c}}$ the 
growth of order begins.
The slower the cooling the more ordered is the system at the end of the cooling
, i.e., at $T=0$ (see Fig.~\ref{ratek}).
To quantify the zero-temperature order we measure the excess energy 
$\delta E$ at $T=0$ and the result is shown in Fig.~\ref{logratek}.

Using the relation (\ref{2}) this data suggests that asymptotically 
$l\sim r^{-1/2}$.
Such a relation, which indicates that the growth of order is relatively fast, 
holds for the two-spin Ising model~\cite{CORNELL} and also for the SS 
model~\cite{SHORE}.
However, this conclusion is based on the validity of the relation (\ref{2}) 
and, similarly to the four-spin model~\cite{LIPDES}, we want to argue that this
relation does not hold.
Our argument refers to the following property of all gonihedric models: the 
energy of cubic-like excitations scales as their linear size 
$L$~\cite{COMM1}.
Let us recall that in two-spin Ising model, this energy scales as the area of 
the excitation ($\sim L^2$).
Provided that the final configuration is composed of such domains of the 
size $L$ and using a simple dimensional argument~\cite{LIPDES2} we obtain
\begin{equation}
l \sim \frac{1}{(\delta E)^{1/2}}.
\label{3}
\end{equation}
A direct confirmation of the assumption about the structure of the 
configuration at the of the cooling process comes from visual inspection.
In Fig.~\ref{configk} we can see an example of single-layer configuration.
One can clearly see  cubic-like (i.e., non-rough) domains whose energy scales
linearly with their size.
Using the relation (\ref{3}) Fig.~\ref{logratek} shows that the 
zero-temperature characteristic length scales as $r^{-1/4}$ which is much slower 
than in the two-spin Ising model but faster than in the four-spin model.

In this section we used three independent techniques to probe the dynamics
of the gonihedric model in the case of continuous thermodynamic transition 
$k=2$.
Domain coarsening suggest that the model has a slow dynamics up to, at least, 
the temperature $T=1.9$ ($T_{{\rm c}}\sim 2.35$).
An analysis of the size dependence of the characteristic time $\tau$ suggests 
that cubic-like domains remain non-rough at least up to the temperature $T=2.1$
Thus, a slow-dynamics regime is most likely extended up this temperature.
Since this is very close to the critical point is it not unlikely that a 
slow-dynamics regime actually covers the whole low-temperature phase.
The behaviour of the model under cooling confirms such a scenario: if there 
would be a certain temperature $T_0<T_{{\rm c}}$ such that for 
$T_0<T<T_{{\rm c}}$ the dynamics would be fast then for the slow cooling the 
growth of order would be dominated by the time spent in this temperature 
interval and we would have $l\sim r^{-1/2}$.
Such a scenario takes place in the SS model~\cite{SHORE}.
The growth of order in our $k=2$ model is much slower $l \sim r^{-1/4}$ and 
excludes the existence of such a temperature $T_0$ (unless it is very close to 
$T_{{\rm c}}$ and our simulations are not sufficient to detect the true 
asymptotic behaviour).
Let us also notice that Shore {\it et al.} also analyzed certain SOS model for
which (by necessity) $T_0=T_{{\rm c}}$.
Using some scaling arguments they have shown that for this model one should 
have $l\sim r^{-1/4}$, which is in agreement with our numerical result.
\section{Off-gonihedric Ising model}
The gonihedric case corresponds to a certain choice of interaction constants
in the Hamiltonian (\ref{1}).
As we have noted this choice has important implications: the ground state is strongly degenerate
and energy of excitations scale as their linear size and not as their area.
In the present section we examine what is going on when the interaction 
constants of model (\ref{1}) deviate from the gonihedric case~\cite{CAPPI}.

As a particular example we choose: $J_1=6,\ J_2=-1,\ J_4=1/2$, which differs
from the gonihedric case $k=2$ by a modified nearest-neighbour coupling $J_1$.
Such a model has a double-degenerate (ferromagnetic) ground state  and our 
rough estimation of the critical temperature is $T_{{\rm c}}\sim 12.5$.

To examine the dynamics of this model we used the same techniques as described
in the previous section.
First, we examined the coarsening behaviour of this model.
At low temperature (up to $T\sim 3.0$) we observed a very slow decrease of the
energy toward the ground state value.
Our data, which we do not present suggests that for such temperatures $l$ most 
likely increases logarithmically with time.
However, above this temperature dynamics becomes much faster and presumably 
the characteristic length increases as $l\sim t^{1/2}$.

Such behaviour is confirmed by the measurements of the characteristic time 
$\tau$ defined in the same way as in the previous section.
Results of simulations are shown in Fig.~\ref{tauoff}.
They indicate that for temperature $T=6$ and 8 the characteristic time $\tau$
increases as $L^2$ similarly to the SS model above the corner rounding 
transition.

The above results indicate that the dynamical behaviour of the model is 
very similar to the SS model.
Namely, in the low temperature regime the model has a slow dynamics and rapidly
(faster than $L^2$) increasing characteristic time $\tau$.
However, within the ordered phase (i.e., for $T<T_{{\rm c}} \sim 12.5$ there
is also a high temperature regime where dynamics is much faster.
Presumably, in this regime the dynamics is similar to other nonconservative 
systems with scalar order parameter~\cite{BRAY}.
Similarity between the model examined in this section and the SS model is in
our opinion related with the structure of the ground state: both models have 
only double degenerate (ferromagnetic) ground states.
It is well known in the statistical mechanics that degeneracy of the ground 
state plays a very important role in determining the thermodynamic behaviour
of models (e.g., critical behaviour).
It is probable that this is also 
an important factor for determining the dynamical 
behaviour of models.
\section{Nature of the glassy phase}
In the present paper we have examined the dynamics of models described by the
Hamiltonian~(\ref{1}).
Depending on the interaction constants we can distinguish three types of 
behaviour which are schematically shown in Fig.~\ref{diagramy}.

\noindent
{\it (a) Double-degenerate ground state}\\
This, the most typical situation,  appears in the off-gonihedric model studied 
in previous section and also in the SS model whose dynamics has been already
examined in great details~\cite{SHORE}.
The dynamics of the model a low temperatures ($T<T_{{\rm c}}$) has two regimes
separated by a certain temperature $T_{{\rm cr}}$.
For $T<T_{{\rm cr}}$ the model has slow dynamics with most likely 
logarithmically increasing characteristic length $l$.
Such behaviour is related with the fact that at such temperature the model 
is below the corner-roughening transition~\cite{WORTIS}.
As a result, an evolving quench develops complicated structures of 
cubic-like excitations which are very stable and effectively block further
coarsening dynamics~\cite{COMM2}.
At $T=T_{{\rm cr}}$ the model undergoes the corner rounding transition and the 
blocking mechanism is no longer effective.
As a result the fast (standard) dynamics is restored.

\noindent
{\it (b) Gonihedric case with continuous transition}\\
In this case the entire low-temperature phase has  slow dynamics, whose 
origin is similar to the case (a).
Namely, a quench develops cubic-like structures which block further 
coarsening dynamics.
The degeneracy of the ground state seems to be the most important difference 
between this and the off-gonihedric case.
Thus, we relate the disappearance of the corner-rounding transition (or maybe 
its overlap with $T_{{\rm c}}$) with the infinite degeneracy of the ground state.

\noindent
{\it (c) Four-spin model}\\
As we already mentioned, the four-spin model undergoes a dynamic
transition which exhibits a lot of similarities with the glassy transition.
It might be interesting to examine whether such a behaviour appears only at
$k=0$ or persists also for some other (small) values of $k$.

\noindent
{\it Glassy transition: loss of surface tension}\\
Why does this transition exist in the four-spin case and not in the other 
cases?
It was already suggested that the difference between the four-spin model 
and the SS model is related to the degeneracy of the ground 
state~\cite{LIPDES2}.
However, the above analysis of the gonihedric model with $k=2$ shows that the 
infinite degeneracy of the ground state is not sufficient for the model to 
have a glassy transition (of course, we limit our analysis to models which 
can generate diverging energy barriers and thus have slow coarsening dynamics).
Why does the gonihedric case $k=2$ differ from the $k=0$ case (i.e. the 
four-spin model)?
Both models have strongly degenerate ground state.
The degeneracy equals $2^{3L}$ for $k=0$ and $2^L$ for $k=2$.
Since in both cases degeneracy increase exponentially with the linear system 
size this difference does not seem important.
In our opinion, however, the difference in the dynamical behaviour is 
related with the ground state structure of these models.
As we already mentioned, for $k=0$ the flipping planes, which generate 
various ground-state configurations might cross.
As a result, in addition to ferromagnetic-like configurations we obtain 
antiferromagnetic-like ones.
For the $k\neq 0$ such crossings are not allowed and only 
ferromagnetic-like configurations are possible (we consider lamellar 
configurations also as ferromagnetic-like).

This difference has important implications.
Let us note that for the four-spin model in addition to tensionless domain 
walls, which appear for example when a cubic ferromagnetic 'up' domain is 
surrounded by 'down' one, there are tensionful ones too.
As an example of such a domain wall we can consider an antiferromagnetic 
domain surrounded by ferromagnetic one~\cite{LIPDES2}.
At first sight it does not seem to be much different from the $k=2$ case.
Indeed, when one considers a lamellar configuration where successive layers 
are of opposite sign (see Fig.~\ref{lamellar}) which is surrounded by a 
ferromagnetic domain than the excess energy of 
such a configuration scales as the area of the wall (i.e., the domain wall is 
tensionful).
There is, however, an important question: do such configurations affect 
coarsening dynamics or, in other words, are they spontaneously generated in 
sufficient amounts?
In our opinion the answer to this question is negative.
Our first argument comes simply from the visual inspection of the snapshot 
configuration.
In Fig.~\ref{configk} one can see relatively large ferromagnetic-like 
domains but there is no indication of lamellar ones.
The second argument comes from the Cluster Variational Method 
calculations~\cite{CIRILLO} by Cirillo {\it et al.} who have shown that the 
lamellar structures are equivalent (i.e., of the same energy) to the 
ferromagnetic one but only for the ground state.
At non-zero temperatures they are always metastable.
These arguments show why such configurations are not 
spontaneously generated during the evolution of the quench.
They imply that for $k=2$ the dominant domain walls which exist at the 
late-time evolution of the quench are tensionless.

On the other hand, for the four-spin model, antiferromagnetic structures are 
fully equivalent to the ferromagnetic ones.
In the liquid phase both ferromagnetic and antiferromagnetic domains
are intertwined and form  very complicated structures.
As already noticed~\cite{LIPDES2}, tensionful domains usually have
lowest energy barriers and the system can relatively easily remove the 
interior domains.
On the contrary, tensionless domain walls have large energy barriers and their
dynamics is much slower.
It means that in the liquid phase dynamics is dominated by dynamics of
tensionful domains and thus resembles the dynamics of two-spin Ising models.

It is in our opinion very likely that upon lowering the temperature the system 
will undergo a phase transition which will eliminate tensionful domain walls.
Below that transition the energy of the system would be located mainly in 
tensionless domain walls.
It means that at this transition the antiferromagnetic-ferromagnetic symmetry 
of the model would be spontaneously broken.
In other words, at this transition the system selects a dominant type of 
domains, whether ferromagnetic or antiferromagnetic.
This, rather novel, type of symmetry breaking is an essential ingredient
of the transition which we tentatively identify as the glassy transition.

Finally, let us note that the idea that the glassy phase consists of a complicated mixture of tensionless domain walls appeared recently in the context of spin glasses~\cite{HOUDAYER}.
It suggests that, at least at the geometrical level, spin glasses and ordinary glasses might exhibit a lot of similarities.
Their further explorations is, however, left as a future problem.
\acknowledgements
{This work was partially supported by
the EC IHP network
``Discrete Random Geometries: From Solid State Physics to Quantum Gravity''
{\it HPRN-CT-1999-000161} and the
ESF network ``Geometry and Disorder: From Membranes to Quantum Gravity''.
The work of DJ and DE was also partially supported by an Acciones Integradas
grant.}

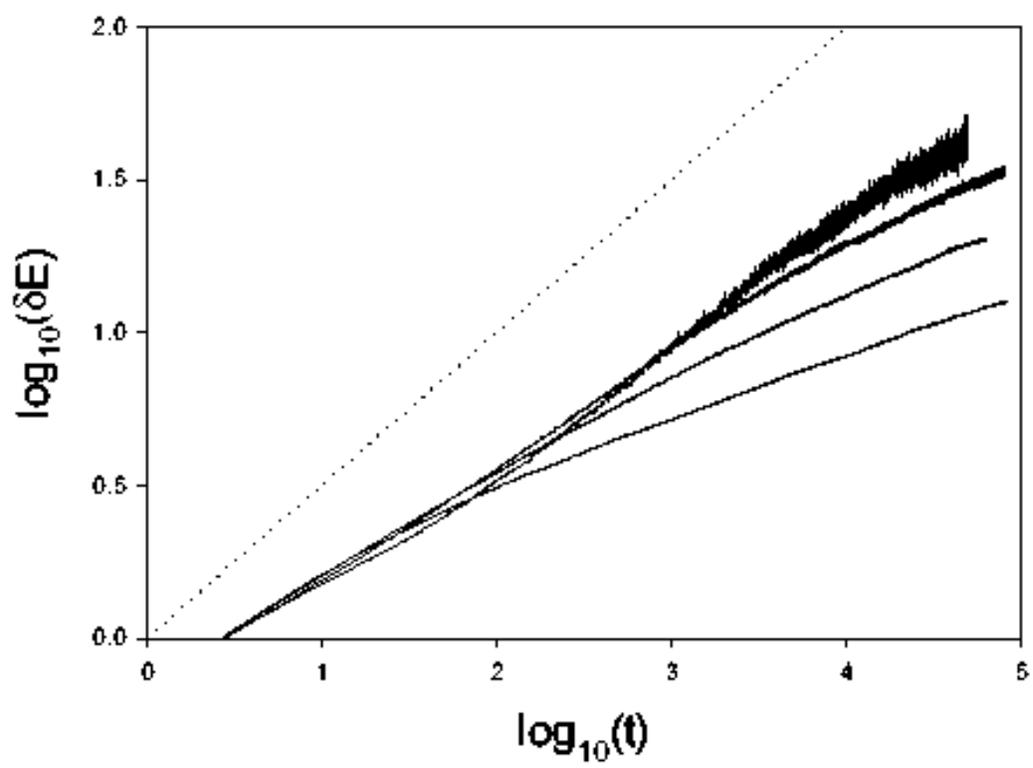
\begin{figure}
\caption{
The excess energy $\delta E$ as a function of $t$ in the log-log scale.
Simulations were made for the system size $L=100$ and $T=1.0$, 1.3, 1.6 and 1.9 (from bottom to the top).
The dotted line has a slope 0.5.
}
\label{f1} 
\end{figure}
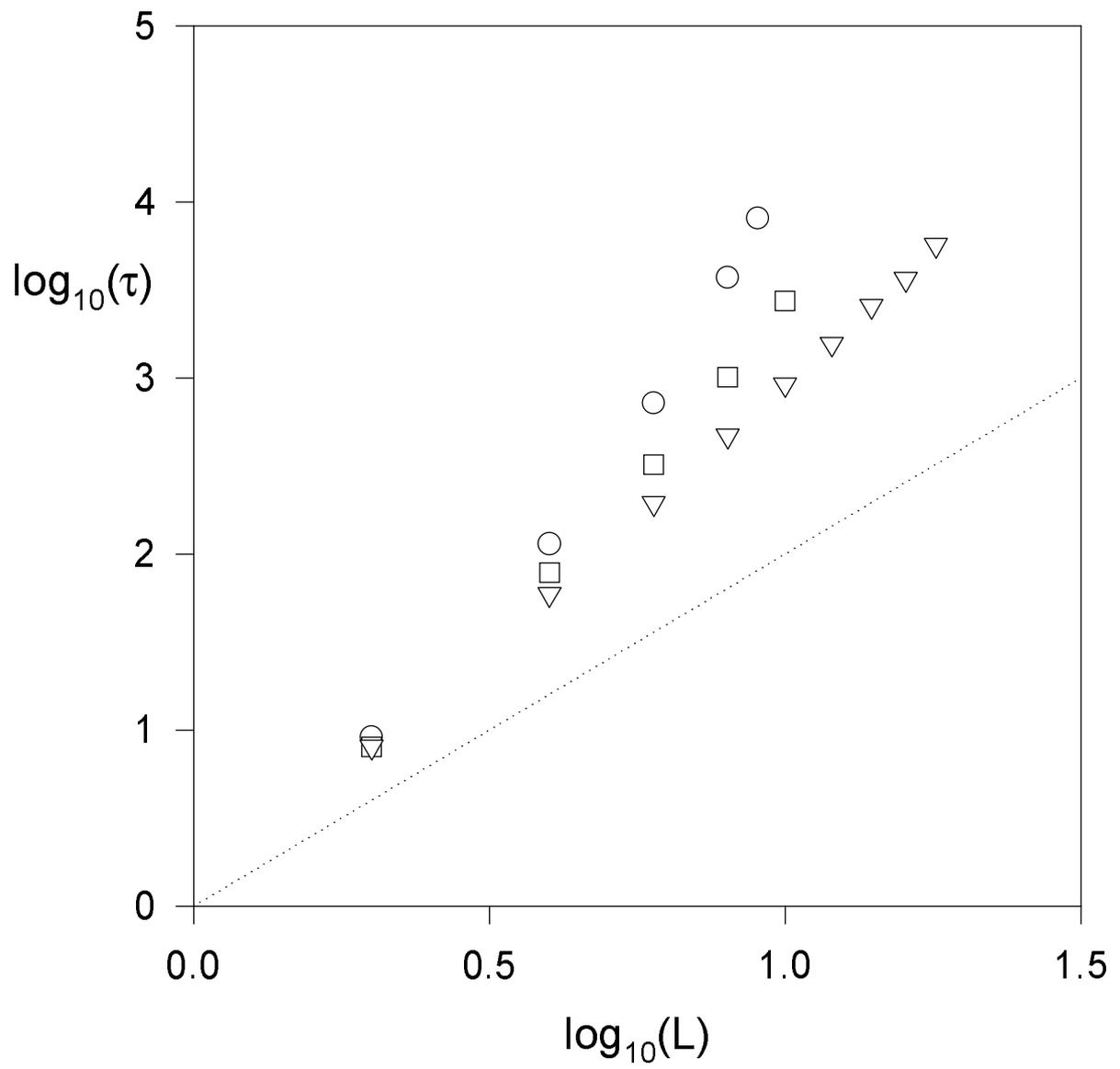
\begin{figure}
\caption{
The characterisitc time $\tau$ needed for the inversion of the cubic excitation
of the size $L$ as a function of $L$ for the gonihedric case ($k=2$).
Simulations were made for $T=1.9$ ($\circ$), 2.1 ($\Box$) and 2.3 ($\nabla$).
The dotted line has a slope corresponding to $\tau \sim L^2$.
}
\label{f2} 
\end{figure}
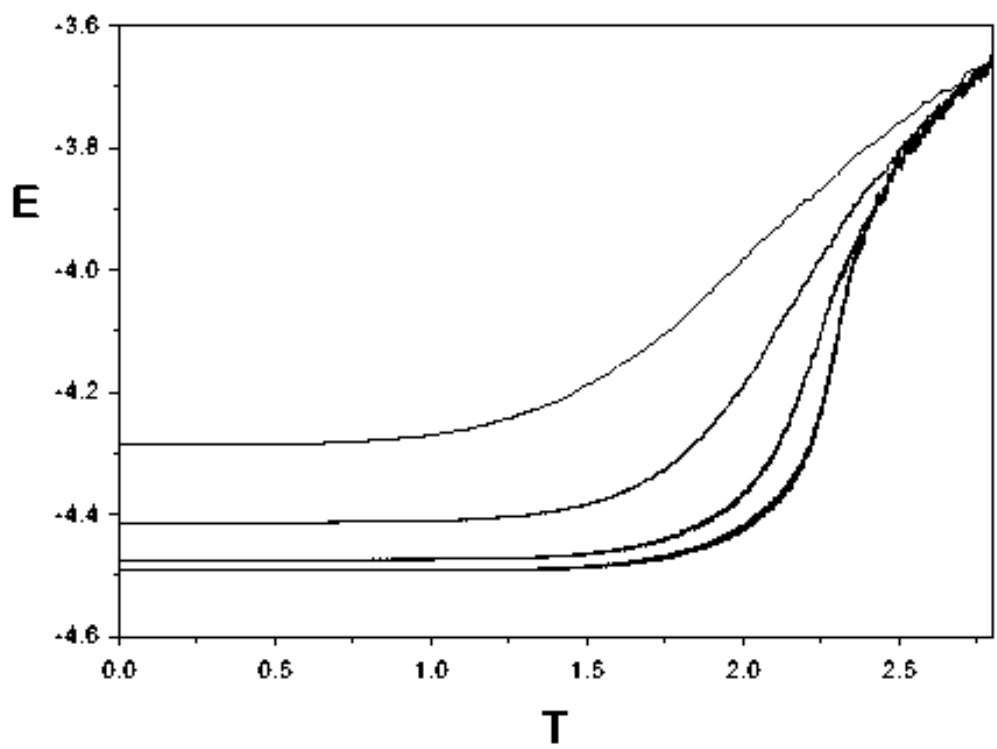
\begin{figure}
\caption{The energy $E$ as a function of temperature for (from the top) $r=$ 0.02, 
0.002, 00002 and 0.00002.
The ground-state energy equals (for $k=2$) -4.5.
}
\label{ratek} 
\end{figure}
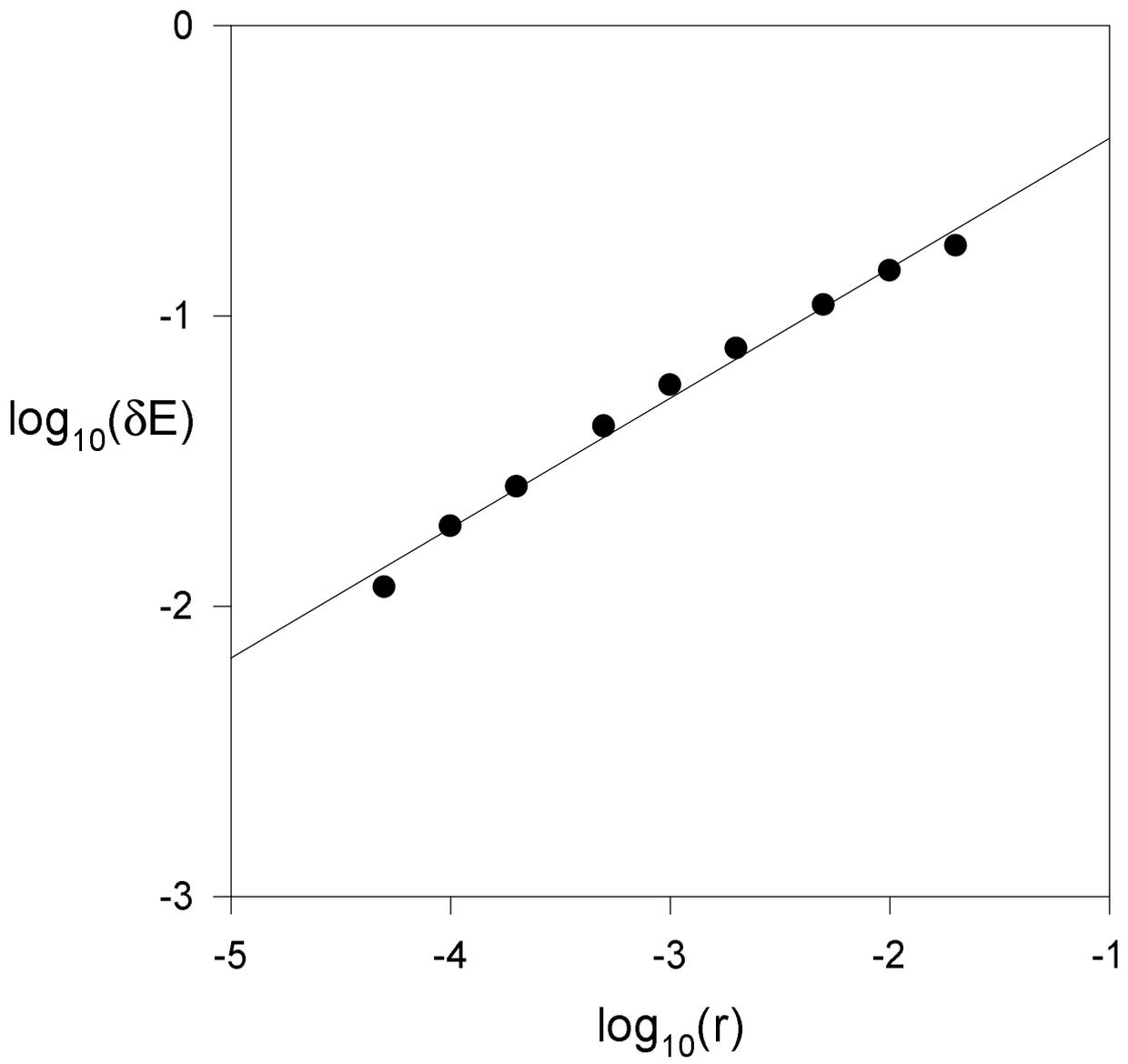
\begin{figure}
\caption{The excess energy $\delta E$ as a function of $r$ in the log-log
scale.
The dotted line has slope 0.45.}
\label{logratek} 
\end{figure}
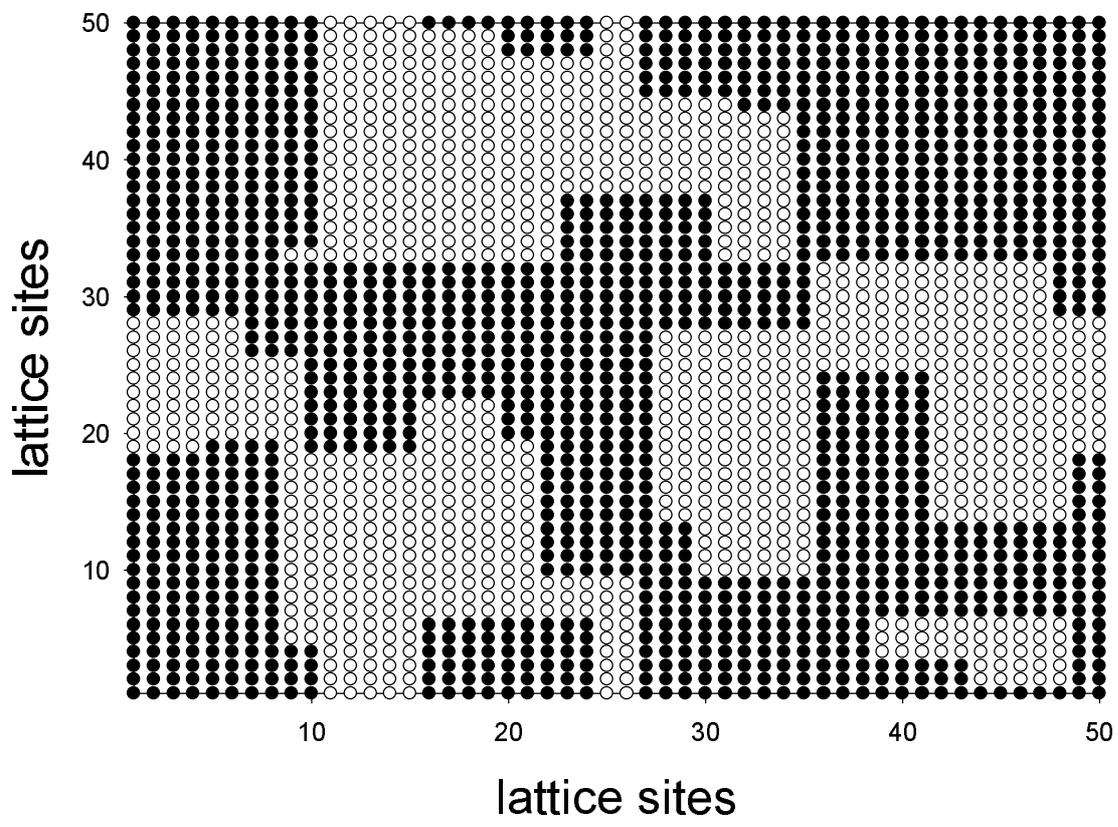
\begin{figure}
\caption{
A single-layer snapshot of a Monte Carlo configuration at the end of the 
cooling process.
Simulations were made for $L=50$ and $r=0.001$.
}
\label{configk} 
\end{figure}
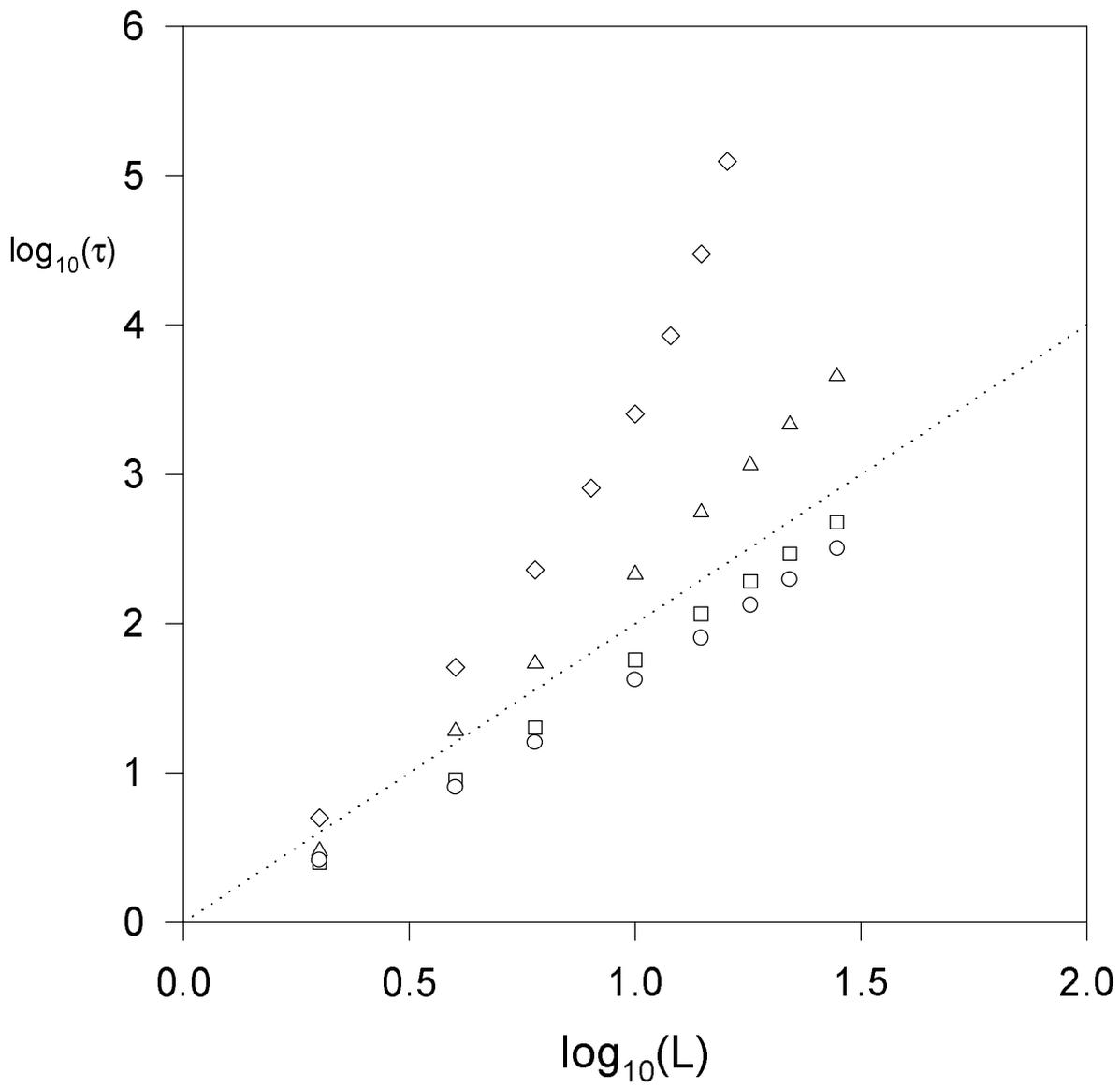
\begin{figure}
\caption{
The characteristic time $\tau$ needed for the inversion of the cubic excitation
of the size $L$ as a function of $L$ for the off-gonihedric case.
Simulations were made for $T=2$ ($\diamondsuit$), 3 ($\triangle$), 6 ($\Box$) and 8 ($\circ$).
The dotted line has a slope corresponding to $\tau \sim L^2$.
}
\label{tauoff} 
\end{figure}
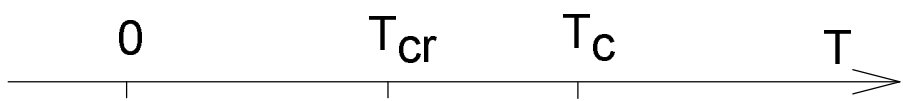
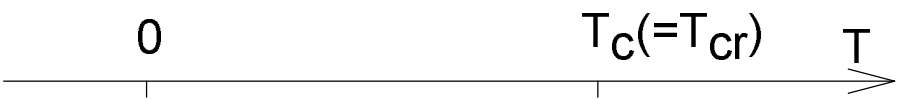
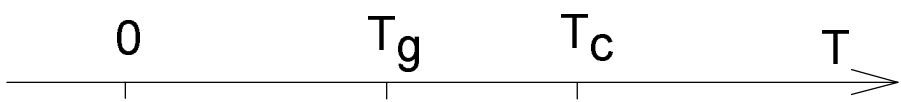
\begin{figure}
\caption{
Three types of dynamical behaviour found for model~(\protect\ref{1}).
(a) Double-degenerate case (SS model); (b) the gonihedric model with continuous
thermodynamic phase transition; (c) the four-spin model (and possibly the 
gonihedric model with discontinuous thermodynamic phase transition).
}
\label{diagramy} 
\end{figure}
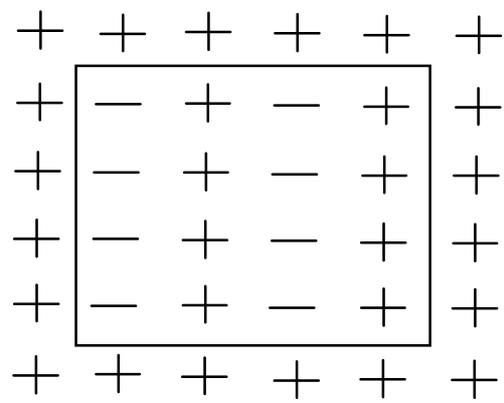
\begin{figure}
\caption{
A two dimensinal (4x4) lamellar domain surrounded by the 
ferromagnetic domain.
The excess energy comes only from horizontal boundaries of domain wall and 
thus scales as the size of the interior domain.
In the three-dimensional case the excess energy would scale as an area of
interior domain.
}
\label{lamellar} 
\end{figure}

\begin {references}
\bibitem{COMM1} Actually, such a scaling of energy of excitations was the main
motivation to construct such models~\cite{SAVVIDY}
\bibitem{COMM2} Coarsening proceeds, basically by removal of interior 
domains.
However, below the corner rounding transition the removal of cubic-like domains
requires climbing energy barriers which increases linearly with the size of 
domains.
\end{references}
\end {document}